# Three-Dimensional Electrode Integration with Microwave Sensors for Precise Microparticle Detection in Microfluidics

Yagmur Ceren Alatas[1,2], Uzay Tefek[1,2], Berk Kucukoglu[1,2], Naz Bardakci[1,2], Sayedus Salehin[1,2], M. Selim Hanay[1,2]

[1] Department of Mechanical Engineering, Bilkent University, Ankara, 06800 TURKEY

[2] UNAM – Institute of Materials Science and Nanotechnology, Bilkent University, Ankara, 06800 TURKEY

**ABSTRACT**

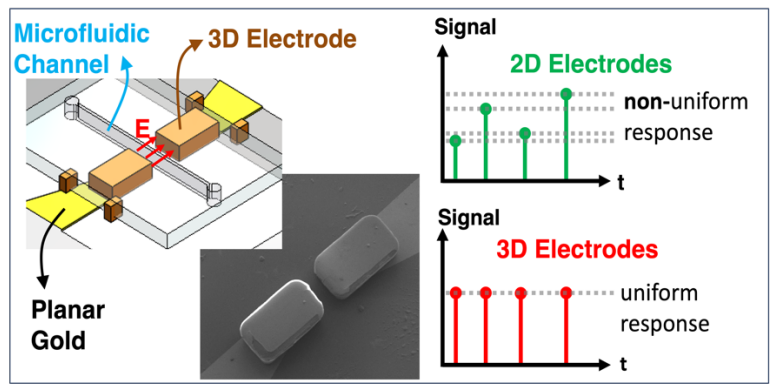

Microwave sensors integrated with microfluidic platforms can provide the size and permittivity of single cells and microparticles. Amongst the microwave sensor topologies, the planar arrangement of electrodes is a popular choice owing to the ease of fabrication. Unfortunately, planar electrodes generate a non-uniform electric field which causes the responsivity of the sensor to depend on the vertical position of a microparticle in the microfluidic channel. To overcome this problem, we fabricated three-dimensional (3D) electrodes at the coplanar sensing region of an underlying microwave resonator. The 3D electrodes are based on SU8 polymer which is then metallized by sputter coating. With this system, we readily characterized a mixture composed of 12 μm and 20 μm polystyrene particles and demonstrated separation without any position-related calibration. The ratio of the electronic response of the two particle types is approximately equal to the ratio of the particle volumes, which indicates the generation of a uniform electric field at the sensing region. The current work obviates the need for using multiple coplanar electrodes and extensive processing of the data for the calibration of particle height in a microfluidic channel: as such, it enables the fabrication of more sophisticated microwave resonators for environmental and biological applications.

# 1. INTRODUCTION

Electronic sensors integrated with microfluidics constitute a powerful platform for the label-free analysis of cells and microparticles since it can reach single-particle resolution at high-throughput.[1] When a microparticle passes through the sensing region of a sensor, it modifies the electric field distribution and induces a shift in the impedance of the sensor. Conventionally, electronic sensors have been designed to work at low frequencies where the conductivity of the solute ions plays an important role. For instance, in resistive pulse sensing,[2] an ionic current is sustained by the solute ions acting as charge carriers: this current gets partially obstructed by a dielectric microparticle partially passing between the two electrodes. As a result, the dielectric particle causes a change in the ionic current proportional to its geometric size. Resistive pulse sensing has been used in a broad range of applications from cell counting in Coulter devices, to DNA sequencing in nanopore devices.[3]

In contrast to resistive pulse sensing, which works at relatively low alternating current (AC) frequencies, an emerging approach is to use capacitive sensors operating at higher frequencies, e.g. microwave band,[4-22] where the ion motion can no longer follow the rapidly changing electric field and ionic current is therefore bypassed. At the microwave band, microparticles induce a shift in the capacitance of the sensor, proportional to the square of the electric field at the sensing region. This capacitance shift can be measured with high accuracy by using resonance-based detection. The change in capacitance translates into a change in the phase, amplitude or resonance frequency of the resonator which can be measured by phase-sensitive detection.[4-6,23] One critical advantage of capacitive sensing is that the signal contains information not only on the particle size, but also on its permittivity. This way, with a suitable sensor arrangement, it becomes possible to distinguish different microparticles from each other based on permittivity.[24]

In microfluidics-integrated electronic sensors, electrodes can be typically configured in three different ways: in a coplanar arrangement, as a parallel-plate structure with metal surfaces at the floor and ceiling of the microfluidic channel, or 3D electrodes at the sidewalls of the channel. Among these configurations, the fabrication of coplanar electrodes is relatively simple, however the electric field generated in the sensing region is non-uniform: its magnitude decays with the distance from the electrodes along the altitude of the microfluidic channel — which causes the electrical signal to depend on the particle trajectory along the channel. Although there are emerging techniques to tackle this problem, e.g. by conducting multiple electronic measurements on the same microparticle at RF[25-27] and microwave[24] band, sensors with coplanar electrodes suffer from the uncertainty in particle location within the channel.

Compared to coplanar sensors, parallel-plate and 3D electrodes require more steps in microfabrication, yet a uniform electric field can be generated between the electrodes, mitigating the positional dependency of the particle signal. In literature, 3D electrodes have been reported in the form of solid or conductive liquid form. While conductive liquids (as electrolytes or liquid metals) can be readily filled into access channels to obtain a 3D filling, solid electrodes with height on the order of tens of microns require more involved metallization procedures. Detailed discussion on liquid electrodes and other electrode configurations can be found in recent references.[28] In particular, Wang et al. designed and fabricated thick electrodes (30-40μm) at the sidewalls of SU8 microfluidic channel with electroplating. This device was used for magnetohydrodynamics and a Dielectrophoresis (DEP) applications.[29] Leroy et al. have also utilized electroplating to obtain ~15 μm gold electrodes on fused silica substrate. These electrodes were used for capacitive detection of trapped microparticles and cells in broadband and resonant architectures at microwave frequency range inside a microfluidic system.[30] Palego et al. have adopted BiCMOS process to integrate microfluidic channel and electrodes (maximum thickness of 9 μm) into the back-end-of-line (BEOL) for electrical cell trapping and sensing with microwave intermodulation.[31] Iliescu et al. fabricated a DEP microfluidic chip with highly doped, thick silicon electrodes at the sidewalls of the microfluidic channel.[32] They used well-established techniques for silicon etching for low-frequency electronic operation, however the use of semiconductors for microwave sensors typically results in resonators with high dissipation. Blanket electrodes were used recently to obtain uniform electric field.[33]

High aspect ratio conductive structures can also be fabricated through methods outside cleanroom procedures. Mechanical micromachining can be used to manufacture electrodes (with heights of typically ~100 μm) which can then be placed at microfluidic channel walls.[34] Although this method yields reusable electrodes, and it can yield electrodes as thin as 20 μm as demonstrated in DEP applications,[35] it requires the manual placement of electrodes into channel sidewalls. Electrospray ionization (ESI) was also reported as a method for obtaining high-aspect ratio, conductive traces on surfaces patterned with photoresist. A wide range of particles such as metal, semiconductor, and dielectric can be electrospray patterned on surfaces[36,37] even on active sensor surfaces.[38] Min & Kim used 3D printer direct writing to metallize dielectric structures with 50 μm thickness through coating custom-made silver paste solution. Silver paste coated electrodes were used for droplet manipulation applications.[39] Similar approaches in 3D printing, such as aerosol jet printing, are being increasingly used for defining microwave circuits at frequencies exceeding 30 GHz.[40]

Finally, 3D electrodes can also be obtained by modifying soft-lithography workflows, such as by patterning photosensitive polymers such as SU8 and Durimide. Normally, such polymers are used in microfluidic process-flows as templates for the PDMS layer defining microchannels. However, it is also possible to use these polymers as structural materials. In this approach, structures with tens of micrometers thickness can be formed by spin coating the polymer and performing photolithography: however, an additional metallization step is required to form electrodes. Hu *et al.* fabricated a microfluidic chip for cell electrofusion and electroporation using Durimide 7510, a photosensitive polymer precursor, to pattern structures with 25 µm thickness around the microfluidic channel followed by gold sputtering to form electrodes.[41] Due to deformation of Durimide at curing stage, the resultant 3D electrodes had a trapezoidal wall profile instead of the more ideal prismatic shape. SU8, an epoxy-based negative tone photoresist was also utilized in the literature to form 3D microelectrode arrays: Martinez-Duarte *et al.* used pyrolysis process where the SU8 was heated up to $900^0$ C in the absence of oxygen to transform SU8 into carbon to form 3D electrodes for DEP applications.[42] Silver nanoparticle-SU8 composite was also used for 3D electrodes due to its relatively simple fabrication process.[43] Kilchemann *et al.* fabricated metal coated SU8 microelectrodes around a microfluidic channel through sputtering metal on the SU8 structures. With this technique, the thickness of deposited metal can be on the order of nanometers: yet, the deposition profile should be conformal to obtain a uniform electric field.[44] The 3D electrodes fabricated in these studies have been used at low AC frequencies, for trapping cells and analyzing their dielectric properties through electrorotation,[22] so the applicability of the polymer based 3D electrodes at the microwave band has not been demonstrated yet.

In this work, we integrated metal coated SU8 microelectrodes with a microwave resonant sensor. The underlying sensing region is based on coplanar electrodes which are then overlaid with 3D electrodes. The voltage difference between a pair of coplanar electrodes is extended to metal coated SU8 electrodes interfacing each other within the microfluidic channel to form the 3D sensing region: this way a uniform electric field is established through the channel. The fabrication of the 3D electrodes is based on SU8 polymer which is metallized by sputter deposition due to its conformal deposition profile.[44] Compared to our recent work on liquid 3D electrodes integrated with microwave sensors,[28] there is no longer a need for a spacer material between the microfluidic channel and electrode (such as a PDMS boundary to partition the liquid metal electrode from the main microfluidic channel where analyte particles flow). In this case, SU8 microelectrodes directly contact the solution running through the microfluidic channel, boosting the signal-to-noise ratio (SNR) of the sensor. We conducted size classification experiments with 12 µm and 20 µm polystyrene (PS) particles without any post-processing for particle position inside the channel.

## 2. SENSOR DESIGN AND FABRICATION

The sensor is composed of three substructures (Figure 1a): 1) a microstrip line microwave resonator built on a PCB, 2) a separate sensing region containing 3D electrodes connected to underlying coplanar gold electrodes built on a fused silica chip, and 3) a PDMS microfluidic channel passing through the sensing region. The microwave resonator (substructure 1) is connected to the coplanar gold electrodes on the sensing region (substructure 2) via wirebonds. The coplanar gold electrodes, in turn, are connected to the 3D electrodes (Figure 1b) during the microfabrication process. Due to electrical contact between metallized SU8 and underlying coplanar gold electrodes, a uniform electric field can be obtained at the sidewalls of the microfluidic channel (Figure 1c). SEM images of the sensing region at different angles are shown on Figure 1d and 1e to highlight the three-dimensional structure of the electrodes.

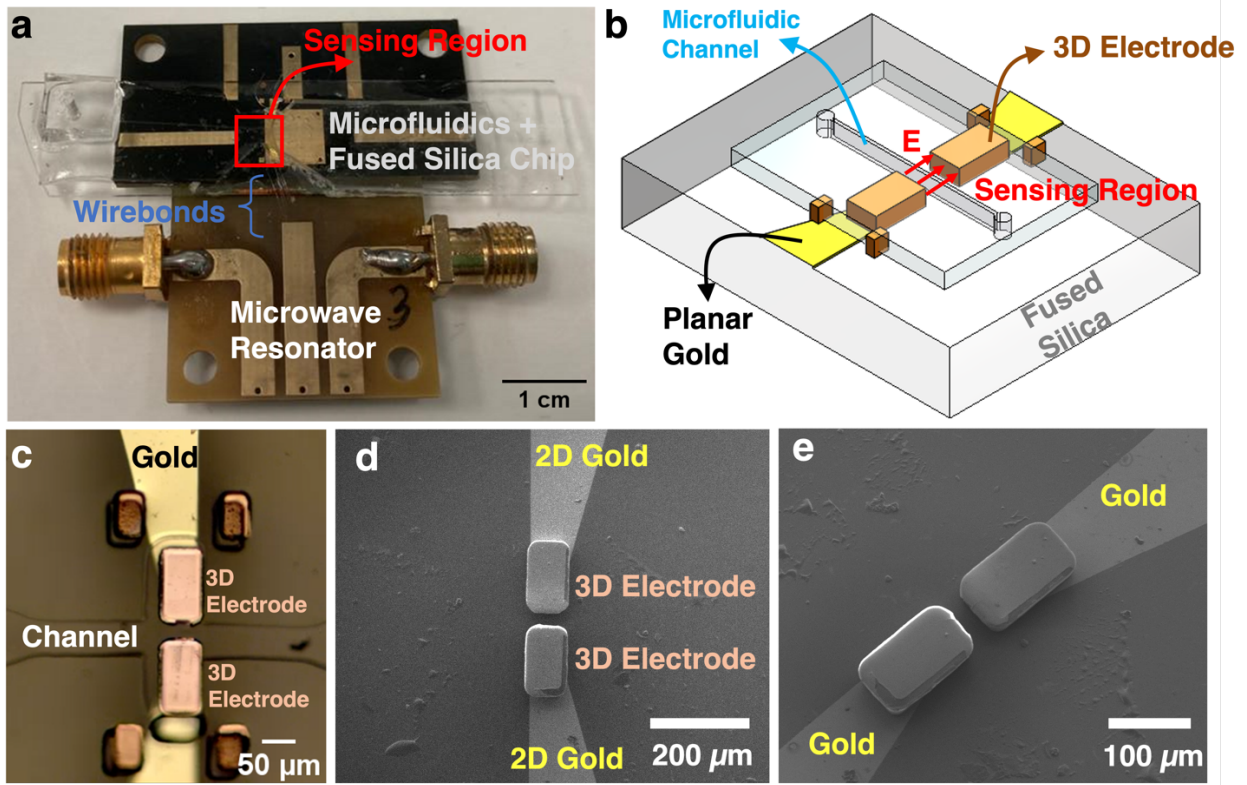

*Figure 1: Microwave sensor at the end of fabrication (a) The microstrip line microwave resonator is fabricated on the lower half of a PCB which carries the fused silica chip with the microfluidic channel on its upper half. The sensing region with a narrow gap between two metal pads and accompanying 3D electrodes are on the fused silica chip. The connections between the sensing region and the microwave resonator are done by wirebonds. (b) Schematic view of the sensing region and the electric field between SU8 microelectrodes (c) Optical microscope image of the sensing region with coplanar gold electrodes and*

*3D copper electrodes. In addition to the two 3D electrodes facing the microfluidic channel, there are four similar structures visible that were used as alignment marks. **(d)** Top view, and **(e)** tilted view of the 3D SU8 microelectrode structure on the coplanar gold electrodes before placing microfluidic channel. To facilitate SEM imaging, the chip was additionally sputtered with a thin layer of Gold/Palladium. Figures a and c are adapted from conference paper [45].*

The fabrication of coplanar and 3D electrodes on the fused silica chip relies on standard cleanroom processes. Fused silica wafer (0.5 mm thickness) was diced into 1.5 cm x 5 cm pieces using a dicing saw. Substrates were cleaned with acetone, IPA, and DI water, and spin coated with AZ5214E photoresist for a final thickness of 1.6 µm. The substrate was exposed to UV light under a photomask and later developed inside AZ400K solution. To fabricate the coplanar electrodes, chromium adhesion layer (10 nm) and gold layer (120 nm) were deposited using thermal evaporation. Coplanar electrodes were obtained after lift-off inside acetone (Figure 2a).

To fabricate the 3D electrodes, first SU8 2050 negative photoresist was spin coated on the substrates for a final thickness of 50µm. The substrate was exposed to UV light under a second photomask which defines micropillars to serve as the base structure of the 3D microelectrodes at the tip of the coplanar electrodes. Micropillars were developed inside SU8 developer solution (Kayakuam) after post exposure bake. The chip was hard baked at 120º C for 40 minutes to improve the adhesion of the resist on the substrate (Figure 2b); this way SU8 micropillars as the structural elements were fabricated. The next step was the metallization of SU8 micropillars. First a layer of AZ5124E photoresist was spin coated on the substrate with the same final thickness as the initial spin coating. Substrate was exposed to UV light under second photomask and developed inside AZ400K-DI Water solution with a 1:4 volume ratio to obtain a photoresist layer for lift-off after the subsequent step. Aluminum or chromium layer (150 nm) was coated on the substrate through DC sputter deposition so that conformal coating was achieved; interior walls of SU8 structures were uniformly coated. During sputter deposition, substrate was attached to a 45º tilted stub to ensure that inner walls of SU8 structures are coated. SU8 microelectrodes were obtained after lift-off inside acetone (Figure 2c). Uniformity and conformality of sputter deposition were monitored using SEM and EDAX.

After fabricating metallized SU8 electrodes, PDMS microfluidic channel was bonded to the substrate after an oxygen plasma treatment. Microchannel mold for PDMS casting was fabricated using SU8 2050 photoresist (on a separate Si wafer). The photoresist was spin coated on a 4-inch wafer for 50 µm final thickness. After consecutive soft baking at 65º C and 95º C, the wafer was patterned under UV light using microfluidic channel photomask. Post exposure bake was done at 65º C and 95º C consecutively and microchannels were developed inside SU8 developer solution. Hard bake was done at 120º C for an hour. PDMS (Sylgard) was mixed (10:1, elastomer base: curing agent) and degassed in a vacuum desiccator. Liquid PDMS was cast on the SU8 mold and cured at 100º C. Solidified PDMS channel block was peeled from the mold and bonded on fused silica substrate after oxygen plasma treatment (Figure 2d) at the last step of cleanroom fabrication. Electrical connections between the fused silica chip with SU8 microelectrodes and microwave resonator PCB were established through wire bonding.

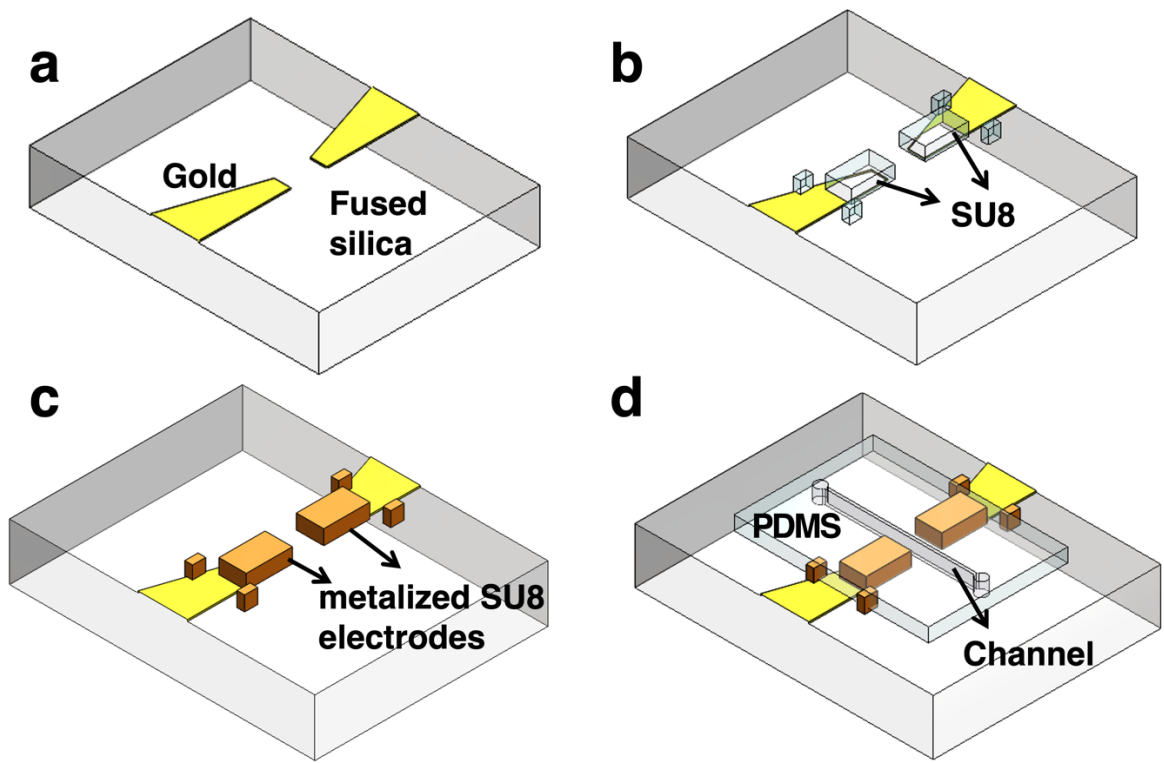

*Figure 2: (a) Coplanar electrodes on fused silica after lift-off. (b) SU8 micropillars patterned at the tip of coplanar electrodes. (c) Metalized SU8 microelectrodes obtained by sputter deposition and lift-off processes. (d) Schematic view of the sensor after a PDMS microfluidic channel was aligned between SU8 microelectrodes and bonded to the fused silica substrate.*

### 3. MEASUREMENT PROCEDURE

The experimental setup consists of three components: (i) Custom-built, phase-sensitive detection circuitry to track the phase and amplitude response of the microwave resonator, (ii) a pressure control system (Fluigent MFCS-EZ) and fluid reservoirs to transport particles into the microfluidic channel, and (iii) an optical microscope to record particle passage events through sensing region and correlate them with corresponding electrical signals. Optical microscope recordings were later used to eliminate particle cluster signals during data analysis. Test solution consisted of the combination of 15 µL of 20-µm (Part number: 74491, Sigma Aldrich) and 15 µL of 12-µm polystyrene (PS) particles (Part number: 88511, Sigma Aldrich) diluted inside 10 mL PBS solution (Biowest). The test solution was pumped into the microfluidic channel during the experiments. Tween 20 (Sigma Aldrich) surfactant was added to the solution (0.1%) as a standard precaution to avoid microfluidic channel clogging.

Before particle sensing experiments, the microwave resonator was initially measured by a vector network analyzer (VNA) to determine the resonance frequency. The resonance frequency of the sensor with 3D electrode was 1.7 GHz (whereas this value was 2.1 GHz for the same device but without the 3D electrodes). During polystyrene particle sensing and size classification experiments, the resonator was connected to a custom-built detection circuitry composed of a signal generator and a lock-in amplifier to achieve phase-sensitive detection.[4-6,23] After transferring the device into the custom-built circuitry and filling the channel with the PBS solution, the resonance frequency of the device was shifted slightly to 1.8 GHz. The phase and amplitude response of the resonator were tracked at this frequency in an open loop measurement configuration as particles were passing through the sensing region between SU8 microelectrodes. The resolution of the lock-in amplifier (phase-sensitive detection) is sufficient to detect small changes in phase and amplitude. A lock-in time constant of 1 ms and sampling rate of 13.4 kSa/s were used in the particle sensing experiments to obtain a high Signal-to-Noise ratio (SNR) at sufficient measurement speeds. Since the resonance frequency of the resonator is above the frequency range of the lock-in amplifier, the reference signal from the lock-in amplifier is up-converted before entering the resonator and the signal from the resonator is down-converted before signals are digitally processed. The details of the circuitry are detailed in the previous work.[23,24] Instantaneous changes in capacitance between SU8 microelectrodes due to single microparticles were expected to be proportional to their volume and Clausius-Mossotti factor. Capacitance change was reflected as a spike-shaped event in phase and amplitude responses.

## 4. RESULTS

Figure 3 shows amplitude shifts of 20 μm PS particles, obtained with 3D SU8 electrodes and coplanar-only electrodes. In the case of the planar sensor, the microwave resonator architecture was based on a Split Ring Resonator with a sensing gap of 20 μm between the 100 nm thick gold electrodes. For both sensors, the same custom measurement circuitry was used. Although the particles were of uniform size, the signal from the sensor with coplanar electrodes (Figure 3b) varied significantly because of positional dependency: yet, the SU8 microelectrode resonator signals consistently yield similar values, independent of the particle trajectory through the sensing region.

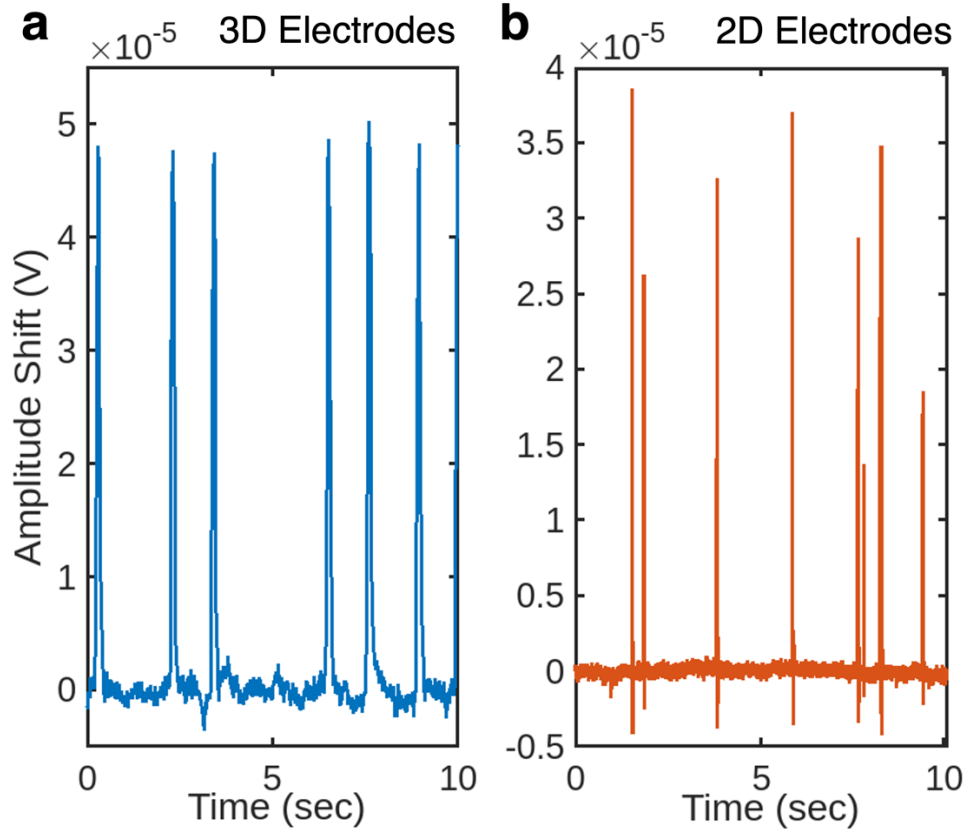

*Figure 3: Amplitude shifts induced by 20µm PS particles with 3D and 2D sensing electrodes. (a) When 3D electrodes are used, as described in this work, the events are consistently of similar size. (b) When a sensor with 2D sensing electrodes is used, there is a large variation in amplitudes, even though the particles tested are nominally at the same size. I*

We further used SU8 microelectrode integrated resonators for size classification of 12 µm and 20 µm polystyrene particles (Figure 4). In the experiments, a solution containing both types of microparticles was passed through the sensor at the same time. Figure 4 shows the phase and amplitude shifts induced by single 12 µm and 20 µm polystyrene microparticles. Two different signal levels are immediately evident reflecting the existence of two different particle populations. The signal levels of the particles are approximately proportional to their volume since they are composed of the same material. The analysis of signal levels is conducted in detail in the next section.

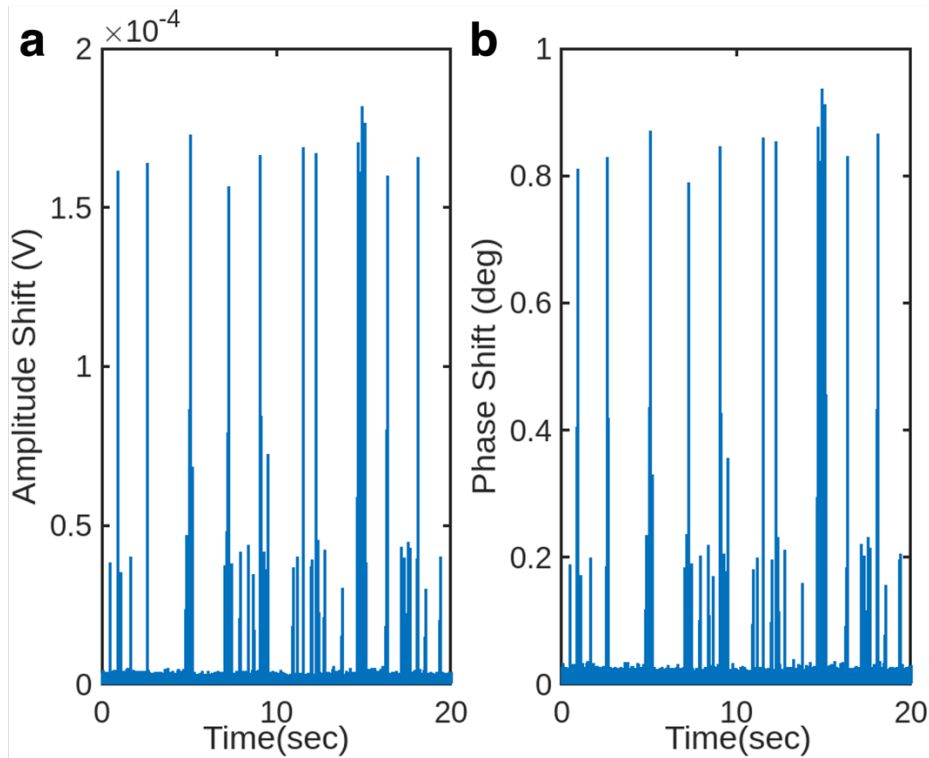

*Figure 4: Amplitude and phase shifts induced by single 12 μm and 20 μm PS particles. Two different height levels are evident corresponding to the larger and smaller particles. Particles of the same size induce similar shifts regardless of their position inside the microfluidic channel.*

## 5. DATA ANALYSIS

We used a MATLAB script to analyze the height of particle-induced shifts in phase and amplitude response of the resonator and correlate with the capacitance change due to each particle. During the experiments, we did not use any control loop to eliminate drift (so as not to decrease the response time of the sensor), so the first step of data analysis was baseline reduction to remove long-term drift in measured signals. Fourier transform of the time domain signal was calculated and power line noise components (including harmonics) were removed using a combination of notch filters.

Particle-induced events were inferred from the amplitude (rather than phase) signal because it had a higher signal-to-noise ratio (SNR). For event detection, the square of the baseline reduced amplitude signal was used to reduce uncertainties. Built-in MATLAB function *findpeaks* was used for the detection of particle-induced events and analysis of their height, width, and location. Prominence threshold was assigned to the peak finder function based on the noise level in control data set which contained no events. As a result of positioning SU8 microelectrodes at the sidewalls of the microfluidic channel, there was direct contact between fluid and electrode, hence SNR was relatively high (~40 for 20 μm particles) and events could be differentiated from noise accurately. The resulting digital peak heights were converted back to voltage and phase values by taking the square root of peak heights.

Since phase and amplitude shifts occur simultaneously, event locations in amplitude signal were used to determine the magnitude of particle-induced shifts in phase signal. To relate the measured values to the particle size, the out-of-phase component (*i.e.,* the Y-quadrature: $Y = R \sin[\theta]$ where R denotes the amplitude, and $\theta$ denotes the phase of the resonator) of the reflected voltage wave was used. The change in the Y-quadrature is proportional to capacitance change due to a particle passage.

$$\Delta Y \propto \frac{\Delta C}{C_0} \quad \ldots (1)$$

where $\Delta C$ denotes the capacitance change induced by the particle, and $C_0$ is the effective capacitance of the sensor.

To express the Y-quadrature shift in terms of measured signals derivative of the constitutive equation can be calculated.

$$\Delta Y = \Delta R \sin(\theta) + R \cos(\theta) \Delta\theta \quad \ldots (2)$$

In addition to sudden changes induced by the particle ($\Delta R$ and $\Delta\theta$), the instantaneous phase ($\theta$) and amplitude (R) values of the resonator were obtained from the data and used in the calculation of the change in the Y-quadrature.

By using the Y-quadrature values, the signals shown in Figure 4 were converted into a histogram shown in Figure 5. The 12 µm and 20 µm particles are clearly separated from each other without any need for calibration of particle altitude.

The clear distinction between the two populations (Figure 5) suggest that a uniform electric field was obtained in the sensing region and the positional dependency of the response was mitigated through the use of 3D microelectrodes at the sidewalls of the microfluidic channel. Microwave signal depends on particle size and permittivity; in this experiment, since all particles were composed of the same material (polystyrene), the signals were expected to be proportional to the volume ratio of the corresponding particles. The ratio of the mean values of Y-quadrature shifts for 20 µm vs 12 µm was 4.4 experimentally, – whereas the expected volume ratio of particles is 4.6. The closeness of the two values (4.4 *vs* 4.6) serves as evidence for the degree of uniformity of the generated electric field. The mean, standard deviation, and coefficient of variation of the Y-quadrature shift are shown in Table 1.

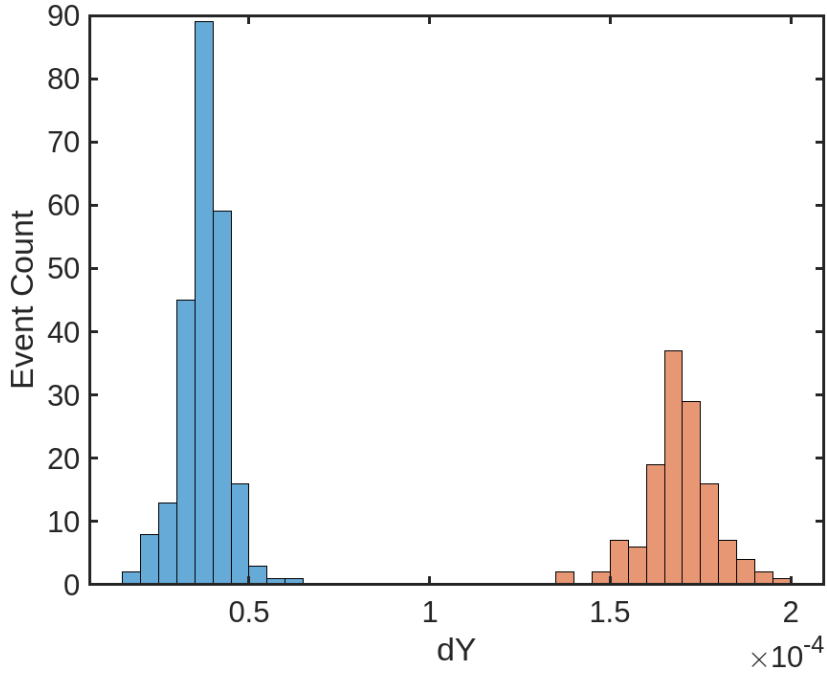

*Figure 5:* *The histogram shows the shift in Y-quadrature induced by 12 µm and 20 µm polystyrene particles. Y-quadrature shift (dY) reflects the normalized capacitance shift due to particles. During the experiment, 369 single particles were measured which were readily separated into two populations.*

| Particle Size | Mean | Standard deviation | Coef. of variation | Number of events |
|---|---|---|---|---|
| *20 µm* | $1.7 \times 10^{-4}$ | $9.8 \times 10^{-6}$ | 0.06 | 132 |
| *12 µm* | $3.8 \times 10^{-5}$ | $6.3 \times 10^{-6}$ | 0.17 | 237 |

*Table 1:* *Statistical parameters of Y-quadrature shifts for 20 µm and 12 µm PS particle mixture experiment*

## 6. CONCLUSION

Particle size classification experiments were conducted using microwave resonators integrated with 3D microelectrodes based on metal coated SU8 structures. SU8, a negative, epoxy-based photoresist was used on the sensor because it is possible to obtain a structural layer with tens of microns thickness through spin coating. Sputter deposition was used for metallization to obtain conformal deposition on the SU8 microelectrodes. A uniform electric field can be established inside the microfluidic channel through positioning SU8 microelectrodes at the sidewalls of the channel. This way positional dependency on the microwave signal was eliminated. The signal ratio between the 12 µm and 20 µm microparticles approached the expected ratio of their corresponding volumes serving as further verification for the uniformity of electric field at the sensing region and the utility of the sensor architecture. Compared to our previous work where we used a liquid metal (Galinstan) to create liquid 3D electrodes,[28] the SNR increased by approximately ten-fold owing to the fact that SU8 microelectrodes are positioned inside the microfluidic channel without any spacer material. This sensor structure eliminates the dependency of microwave sensing signals on the position of particle inside the microfluidic channel; the signal only depends on particle size.

**Acknowledgements**

This project has received funding from the European Research Council (ERC) under the European Union's Horizon 2020 research and innovation programme (grant agreement No 758769). We thank Barbaros Cetin, Murat Güre and Hashim Alhmoud for useful discussion.

**References**

1   Spencer, D. C. *et al.* A fast impedance-based antimicrobial susceptibility test. *Nature communications* **11**, 5328 (2020).

2   Hogg, W. R. & Coulter, W. H. Apparatus and method for measuring a dividing particle size of a particulate system. US3557352 (1971).

3   Brown, C. G. & Clarke, J. Nanopore development at Oxford nanopore. *Nature Biotechnology* **34**, 810-811 (2016).

4   Nikolic-Jaric, M. *et al.* Microwave frequency sensor for detection of biological cells in microfluidic channels. *Biomicrofluidics* **3**, 034103 (2009).

5   Ferrier, G. A., Romanuik, S. F., Thomson, D. J., Bridges, G. E. & Freeman, M. R. A microwave interferometric system for simultaneous actuation and detection of single biological cells. *Lab on a Chip* **9**, 3406-3412 (2009).

6   Kelleci, M., Aydogmus, H., Aslanbas, L., Erbil, S. O. & Hanay, M. S. Towards microwave imaging of cells. *Lab on a Chip* **18**, 463-472 (2018).

7   Afshar, S. *et al.* Multi-frequency DEP cytometer employing a microwave sensor for dielectric analysis of single cells. *IEEE Transactions on Microwave Theory and Techniques* **64**, 991-998 (2016).

8   Yang, Y. *et al.* Distinguishing the viability of a single yeast cell with an ultra-sensitive radio frequency sensor. *Lab on a Chip* **10**, 553-555 (2010).

9	Dalmay, C. *et al.* Ultra sensitive biosensor based on impedance spectroscopy at microwave frequencies for cell scale analysis. *Sensors and Actuators A: Physical* **162**, 189-197 (2010).

10	Chen, T. *et al.* in *2013 IEEE MTT-S International Microwave Symposium Digest (MTT).*  1-4 (IEEE).

11	Rowe, D. J. *et al.* Improved split-ring resonator for microfluidic sensing. *IEEE Transactions on Microwave Theory and Techniques* **62**, 689-699 (2014).

12	Zarifi, M. H., Thundat, T. & Daneshmand, M. High resolution microwave microstrip resonator for sensing applications. *Sensors and Actuators A: Physical* **233**, 224-230 (2015).

13	Chien, J.-C. *et al.* A high-throughput flow cytometry-on-a-CMOS platform for single-cell dielectric spectroscopy at microwave frequencies. *Lab on a Chip* **18**, 2065-2076 (2018).

14	Bhat, A., Gwozdz, P. V., Seshadri, A., Hoeft, M. & Blick, R. H. Tank circuit for ultrafast single-particle detection in micropores. *Physical Review Letters* **121**, 078102 (2018).

15	Narang, R. *et al.* Sensitive, real-time and non-intrusive detection of concentration and growth of pathogenic bacteria using microfluidic-microwave ring resonator biosensor. *Scientific reports* **8**, 15807 (2018).

16	Watts, C. *et al.* Microwave Dielectric Sensing of Free-Flowing, Single, Living Cells in Aqueous Suspension. *IEEE Journal of Electromagnetics, RF and Microwaves in Medicine and Biology*  (2019).

17	Secme, A. *et al.* High-Resolution Dielectric Characterization of Single Cells and Microparticles Using Integrated Microfluidic Microwave Sensors. *IEEE Sensors Journal* **23**, 6517-6529 (2023).

18	Mertens, M., Chavoshi, M., Peytral-Rieu, O., Grenier, K. & Schreurs, D. Dielectric Spectroscopy: Revealing the True Colors of Biological Matter. *IEEE Microwave Magazine* **24**, 49-62 (2023).

19	Krivosudský, O., Havelka, D., Chafai, D. E. & Cifra, M. Microfluidic on-chip microwave sensing of the self-assembly state of tubulin. *Sensors and Actuators B: Chemical* **328**, 129068 (2021).

20	Piekarz, I., Sorocki, J., Gorska, S., Wincza, K. & Gruszczynski, S. in *2023 IEEE MTT-S International Microwave Biomedical Conference (IMBioC).*  109-111 (IEEE).

21	Ferguson, C. A., Hwang, J. C., Zhang, Y. & Cheng, X. Single-Cell Classification Based on Population Nucleus Size Combining Microwave Impedance Spectroscopy and Machine Learning. *Sensors* **23**, 1001 (2023).

22	Grzegorzewski, L., Zierold, R. & Blick, R. H. Coupling-Based Sensing with a Microwave Resonator for Single Nanoscale Particles Detection. *IEEE Sensors Journal*  (2023).

23	Secme, A. *et al.* On-chip flow rate sensing via membrane deformation and bistability probed by microwave resonators. *Microfluidics and Nanofluidics* **27**, 28 (2023).

24      Tefek, U., Sari, B., Alhmoud, H. Z. & Hanay, M. S. Permittivity-Based Microparticle Classification by the Integration of Impedance Cytometry and Microwave Resonators. *Advanced Materials*, 2304072 (2023).

25      De Ninno, A. *et al.* Coplanar electrode microfluidic chip enabling accurate sheathless impedance cytometry. *Lab on a Chip* **17**, 1158-1166 (2017).

26      Zhong, J., Liang, M. & Ai, Y. Submicron-precision particle characterization in microfluidic impedance cytometry with double differential electrodes. *Lab on a Chip* **21**, 2869-2880 (2021).

27      Spencer, D. & Morgan, H. High-speed single-cell dielectric spectroscopy. *ACS sensors* **5**, 423-430 (2020).

28      Alatas, Y. C., Tefek, U., Sari, B. & Hanay, M. S. Microwave Resonators Enhanced With 3D Liquid-Metal Electrodes for Microparticle Sensing in Microfluidic Applications. *IEEE Journal of Microwaves*, 1-8 (2023). https://doi.org:10.1109/JMW.2023.3327521

29      Wang, L., Flanagan, L. & Lee, A. P. Side-wall vertical electrodes for lateral field microfluidic applications. *Journal of microelectromechanical systems* **16**, 454-461 (2007).

30      Leroy, J. *et al.* Microfluidic biosensors for microwave dielectric spectroscopy. *Sensors and Actuators A: Physical* **229**, 172-181 (2015).

31      Palego, C. *et al.* in *2016 IEEE MTT-S International Microwave Symposium (IMS).*  1-4 (IEEE).

32      Iliescu, C., Xu, G. L., Samper, V. & Tay, F. E. Fabrication of a dielectrophoretic chip with 3D silicon electrodes. *Journal of Micromechanics and Microengineering* **15**, 494 (2004).

33      Civelekoglu, O., Liu, R., Asmare, N., Arifuzzman, A. & Sarioglu, A. F. Wrap-around sensors for electrical detection of particles in microfluidic channels. *Sensors and Actuators B: Chemical* **375**, 132874 (2023).

34      Zeinali, S., Cetin, B., Oliaei, S. N. B. & Karpat, Y. Fabrication of continuous flow microfluidics device with 3D electrode structures for high throughput DEP applications using mechanical machining. *Electrophoresis* **36**, 1432-1442 (2015).

35      Çetin, B., Kang, Y., Wu, Z. & Li, D. Continuous particle separation by size via AC-dielectrophoresis using a lab-on-a-chip device with 3-D electrodes. *Electrophoresis* **30**, 766-772 (2009).

36      Zhu, Y. & Chiarot, P. R. Directed assembly of nanomaterials using electrospray deposition and substrate-level patterning. *Powder technology* **364**, 845-850 (2020).

37      Kim, H. *et al.* Parallel patterning of nanoparticles via electrodynamic focusing of charged aerosols. *Nature nanotechnology* **1**, 117-121 (2006).

38      Erdogan, R. T. *et al.* Atmospheric pressure mass spectrometry of single viruses and nanoparticles by nanoelectromechanical systems. *Acs Nano* **16**, 3821-3833 (2022).

39   Min, X. & Kim, W. S. Artificial xylem chip: a three-dimensionally printed vertical digital microfluidic platform. *Langmuir* **36**, 14841-14848 (2020).

40   Piekarz, I. *et al.* Application of aerosol jet 3-D printing with conductive and nonconductive inks for manufacturing mm-wave circuits. *IEEE Transactions on Components, Packaging and Manufacturing Technology* **9**, 586-595 (2018).

41   Hu, N., Yang, J., Qian, S., Joo, S. W. & Zheng, X. A cell electrofusion microfluidic device integrated with 3D thin-film microelectrode arrays. *Biomicrofluidics* **5** (2011).

42   Martinez-Duarte, R., Renaud, P. & Madou, M. J. A novel approach to dielectrophoresis using carbon electrodes. *Electrophoresis* **32**, 2385-2392 (2011).

43   Jiguet, S., Bertsch, A., Hofmann, H. & Renaud, P. Conductive SU8 photoresist for microfabrication. *Advanced Functional Materials* **15**, 1511-1516 (2005).

44   Kilchenmann, S. C., Rollo, E., Maoddi, P. & Guiducci, C. Metal-coated SU-8 structures for high-density 3-D microelectrode arrays. *Journal of Microelectromechanical Systems* **25**, 425-431 (2016).